\documentclass{article}
\usepackage{arxiv}
\usepackage[utf8]{inputenc}
\usepackage[T1]{fontenc}

\usepackage{multirow}
\usepackage{graphicx}
\usepackage{authblk}
\usepackage{booktabs}
\usepackage{url}

\newcommand{\micromole}[0]{$\mu$M}
\newcommand{\nanomole}[0]{nM}
\newcommand{\millimole}[0]{mM}
\newcommand{\microgperml}[0]{$\mu$g/ml}
\newcommand{\nanogperml}[0]{ng/ml}

\graphicspath{{./figures/}}

\title{Machine learning to predict developmental neurotoxicity with high-throughput data from 2D bio-engineered tissues}

\author[1,*]{Finn Kuusisto}
\author[2]{Vitor Santos Costa}
\author[1,+]{Zhonggang Hou}
\author[1,3,4]{James Thomson}
\author[5,6]{David Page}
\author[1]{Ron Stewart}
\affil[1]{Morgridge Institute for Research, Regenerative Biology, Madison, WI, USA}
\affil[2]{University of Porto, Department of Computer Science, Porto, Portugal}
\affil[3]{University of Wisconsin, Department of Cell and Regenerative Biology, Madison, WI, USA}
\affil[4]{University of California, Department of Molecular, Cellular, and Developmental Biology, Santa Barbara, CA, USA}
\affil[5]{University of Wisconsin, Department of Biostatistics and Medical Informatics, Madison, WI, USA}
\affil[6]{University of Wisconsin, Department of Computer Sciences, Madison, WI, USA}
\affil[+]{Current Address: University of Michigan, Department of Biological Chemistry, Ann Arbor, MI, USA}

\affil[*]{Email: fkuusisto@morgridge.org}

\begin{document}
\maketitle

\begin{abstract}
There is a growing need for fast and accurate methods for testing developmental neurotoxicity across several chemical exposure sources.
Current approaches, such as \textit{in vivo} animal studies, and assays of animal and human primary cell cultures, suffer from challenges related to time, cost, and applicability to human physiology.
We previously demonstrated success employing machine learning to predict developmental neurotoxicity using gene expression data collected from human 3D tissue models exposed to various compounds.
The 3D model is biologically similar to developing neural structures, but its complexity necessitates extensive expertise and effort to employ.
By instead focusing solely on constructing an assay of developmental neurotoxicity, we propose that a simpler 2D tissue model may prove sufficient.
We thus compare the accuracy of predictive models trained on data from a 2D tissue model with those trained on data from a 3D tissue model, and find the 2D model to be substantially more accurate.
Furthermore, we find the 2D model to be more robust under stringent gene set selection, whereas the 3D model suffers substantial accuracy degradation.
While both approaches have advantages and disadvantages, we propose that our described 2D approach could be a valuable tool for decision makers when prioritizing neurotoxicity screening.
\end{abstract}

\keywords{machine learning, neurotoxicity, tissue model, gene expression}

\section*{Introduction}
The Toxic Substances Control Act (TSCA) lists 84,000 chemicals, almost all of which have not been tested for developmental neurotoxicity \cite{betts2010growing}.
The developing human brain is especially sensitive to toxic exposures \cite{rice2000critical}, and estimated costs of early developmental neurotoxicity exposure are enormous \cite{grandjean2014neurobehavioural,trasande2011reducing}.
Fast, inexpensive, and accurate methods for testing developmental neurotoxicity are thus urgently needed.

Current approaches involve \textit{in vivo} animal studies, and assays of animal and human primary cell and tissue cultures.
These approaches suffer from challenges such as time, cost, availability of primary cells and tissues, and poor applicability to human physiology \cite{judson2014vitro,fabre2014organs,olson2000concordance}.
The result of these difficulties can be observed in part by a decrease in drug approval rates despite increases in research and development spending \cite{hay2014clinical}.
Human pluripotent stem cells can help address these challenges by providing a scalable source of applicable human cells at relatively low cost.

In 2012, the National Institutes of Health (NIH) launched the Microphysiological Systems Program \cite{fabre2014organs}, a collaboration with the Defense Advanced Research Projects Agency (DARPA) and the U.S. Food and Drug Administration (FDA), to develop human tissue chips containing bio-engineered tissue models that mimic human physiology.
The aim of this program is to use these chips to predict the safety and efficacy of candidate drugs.
Within this program, our lab developed 3D constructs of developing human neural tissue from several cell types and trained machine learning models to predict developmental neurotoxicity with gene expression data gathered from these constructs \cite{schwartz2015human}.
The 3D model is capable of accurately identifying compounds that are neurotoxic, and it shows potential for recapitulating relevant biological mechanisms, thus demonstrating potential for further insight.
Nevertheless, this 3D model is necessarily complex and requires extensive expertise and manual effort to construct and employ successfully.
In cases where the goal is only to construct an assay of developmental neurotoxicity, we propose that it may be sufficient to use a simpler 2D tissue model cultured from a single cell type. 

Here, we report results applying off-the-shelf machine learning algorithms to a simpler 2D model of neural tissue.
We run several experiments with varying numbers of chemical exposure lengths and feature selection methods.
We compare the accuracy of learned models between those trained on data from a 2D tissue model and those trained on data from the 3D tissue model. 
Importantly, we find that our accuracy in distinguishing known human developmental neurotoxins from non-toxins is substantially higher when using the 2D model than when using the 3D model.
We describe our data collection and predictive experiments in the next section, follow with a discussion of results, and finish with conclusions and proposals for future work.

\section*{Materials and Methods}

For all of our experiments, we consider a dataset of 45 compounds.
Our outcome of interest is a binary prediction of toxic or non-toxic, and of these 45 compounds, 29 are considered toxic and 16 are considered non-toxic (see Table \ref{tab:compounds}).
These 45 compounds are a subset of the 70 used in prior work \cite{schwartz2015human}.
For our experiments here, we use the same compound concentrations and the same toxic/non-toxic binary labels assigned to each compound from this prior work.

We collect transcriptome-wide gene expression profiles via RNA sequencing (RNA-Seq)  from 2D tissue cultures following exposure to these compounds.
We use the 45 compound subset of data from the prior 3D work \cite{schwartz2015human} as our comparison gene expression data from 3D tissue cultures.
Throughout the paper, we refer to these datasets as the 2D and 3D dataset, respectively.
Both datasets consist of gene expression measurements in transcripts per million reads (TPM) from 19,084 protein-coding genes for every sample.
Each sample thus represents gene expression from either a 2D or 3D tissue sample, a single compound, and a single length of exposure to that compound.

\begin{table}[!ht]
\centering
\caption{Compounds and concentrations.}
\label{tab:compounds}
\begin{tabular}{c l r}
\cmidrule[\heavyrulewidth]{2-3}
& Chemical & Concentration \\ 
\cmidrule{2-3}
\parbox[t]{1mm}{\multirow{29}{*}{\rotatebox[origin=c]{90}{Toxic}}} & Accutane or Isotretinoin             & 0.5\microgperml\\
& Amiodarone                           & 1\microgperml  \\
& Arsenic                              & 5\micromole    \\
& Benzene                              & 50\microgperml \\
& Bioallethrin                         & 10\nanomole    \\
& Bis-I                                & 4\micromole    \\
& Busulfan                             & 50\microgperml \\
& Cadmium                              & 50\micromole   \\
& Carbamazepine                        & 10\microgperml \\
& Cytosine $\beta$-D-arabinofuranoside & 5\micromole    \\
& Dexamethasone                        & 100\micromole  \\
& Diazinon                             & 10\micromole   \\
& Dioxin                               & 3\nanomole     \\
& 5-Fluorouracil                       & 10\micromole   \\
& Hydroxyurea                          & 250\micromole  \\
& 2-Imidazolidinethione                & 0.4\millimole  \\
& K252a                                & 30\nanomole    \\
& Lead acetate                         & 30\micromole   \\
& Maneb                                & 60\micromole   \\
& Monosodium glutamate                 & 150\micromole  \\
& Okadaic acid                         & 3\nanomole     \\
& PD166866                             & 2.5\micromole  \\
& Permethrin                           & 2.5\micromole  \\
& L-phenylalanine                      & 1\millimole    \\
& Propylthiouracil                     & 25\microgperml \\
& (trans) Retinoic acid                & 1.7\micromole  \\
& Thalidomide                          & 100\nanogperml \\
& U0126                                & 10\micromole   \\
& Vincristine                          & 1\nanomole     \\
\cmidrule{2-3}
\parbox[t]{1mm}{\multirow{16}{*}{\rotatebox[origin=c]{90}{Non-Toxic}}} & Acetaminophen   & 10\microgperml \\
& Amoxicillin     & 5\microgperml  \\
& Aspirin         & 20\microgperml \\
& DMSO            & 0.1\%          \\
& Ficoll 400      & 10\micromole   \\
& Fructose        & 10\micromole   \\
& Glucosamine     & 0.4\microgperml\\
& Glycerol        & 10\micromole   \\
& Glyphosate      & 10\micromole   \\
& Ibuprofen       & 20\microgperml \\
& Naproxen sodium & 30\microgperml \\
& PEG 3350        & 10\micromole   \\
& PVP             & 100\microgperml\\
& Saccharin       & 10\micromole   \\
& Sorbitol        & 10\micromole   \\
& Sucrose         & 10\micromole   \\
\cmidrule[\heavyrulewidth]{2-3}
\end{tabular}
\end{table}

To collect gene expression data for our 2D experiments, we seeded neural progenitor cells (NPC) on Matrigel coated plates and started compound treatment on the same day (day 0).
We collected samples at 11 time points of exposure: 1, 2, 3, 4, 6, 8, 10, 15, 21, 27, and 39 days.
We then performed RNA-Seq and calculated gene expression values in transcripts per million reads.
The dataset contains at most one biological sample for each compound and exposure length.
There are missing samples for some exposure lengths of some compounds because of cell death and other experimental factors (see Table \ref{tab:missing}).

\begin{table}[!ht]
\centering
\caption{Missing samples from the 2D tissue culture dataset.}
\label{tab:missing}
\begin{tabular}{c l r}
\cmidrule[\heavyrulewidth]{2-3}
& Chemical & Missing Days \\
\cmidrule{2-3}
\parbox[t]{1mm}{\multirow{11}{*}{\rotatebox[origin=c]{90}{Toxic}}} & Arsenic & 8 - 39 \\
& Busulfan & 8 - 39 \\
& Cadmium & 8 - 39 \\
& Cytosine $\beta$-D-arabinofuranoside & 8 - 39 \\
& 5-Fluorouracil & 8 - 21 \\
& 2-Imidazolidinethione & 21 \\
& Maneb  & 2 - 39 \\
& Okadaic Acid & 8 - 39 \\
& PD166866 & 8 \\
& U0126 & 21 \\
& Vincristine & 8 - 39 \\
\cmidrule{2-3}
\parbox[t]{1mm}{\multirow{10}{*}{\rotatebox[origin=c]{90}{Non-Toxic}}} & Acetaminophen  & 8, 15 \\
& Aspirin & 8 - 39 \\
& Glucosamine & 15 - 39 \\
& Glycerol & 8 - 27 \\
& Ibuprofen & 15 \\
& Naproxen sodium & 6 \\
& PEG 3350  & 15 \\
& Glyphosate  & 8, 15 \\
& Sorbitol  & 8 - 21 \\
& Saccharin  & 8 - 27 \\
\cmidrule[\heavyrulewidth]{2-3}
\end{tabular}
\end{table}

The 3D dataset contains samples at two time points of exposure: 2 and 7 days.
There are two biological samples for each compound at each time point, with no missing samples.

Overall, the goal of our prediction experiments is then to learn a model that can map from gene expression profiles to a binary label of toxic or non-toxic.
We make no a priori assumptions about the common or unique biological effects of the compounds in the expression data.
Thus, we make no attempt to separate or explicitly model different types of toxicity or non-toxicity that each individual compound may elicit.
Similarly, we make no attempt to explicitly model the effects of different compound exposure lengths.
Instead, we simply allow the machine learning algorithms to observe samples from many different compounds at different exposure lengths and find patterns across the gene expression profiles that are associated with toxicity or non-toxicity.
Using this approach, we aim to develop an accurate predictive model that generalizes beyond any single exposure length or type of toxicity.
The primary advantage of this approach is thus its generalizability, but the disadvantage is that analyzing the model to understand any one particular toxic effect becomes more difficult because all individual signals are effectively combined into one broad signal.

We first give more detailed descriptions of our cell culture approach and our sequencing pipeline in the next two subsections.
We then describe and perform three sets of prediction experiments, performing each separately on the 2D and 3D datasets, and compare results.
The first experiment evaluates the ability to predict toxicity by training and testing common machine learning models on samples from a single time point of compound exposure.
Next, we evaluate the ability to predict toxicity by training and testing models using all available time points of compound exposure pooled together.
Finally, we evaluate the same models once more using all time points of exposure but with the addition of feature selection.

\subsection*{Cell culture and neural differentiation}
\label{sec:cell_culture}
For our 2D tissue model, we maintained human H1 embryonic stem (ES) cells in E8 medium \cite{chen2011chemically} on Matrigel (BD Biosciences) coated plates and passaged every 4-5 days with EDTA \cite{beers2012passaging}.
To derive neural progenitor cells (NPC), we split H1 cells using EDTA and immediately treated them with E7 medium (E8 medium without TGF$\beta$-1) +SB431542 (Sigma) (10\micromole).
After another seven days, we mechanically dissociated the formed rosettes from the culture dish and cultured the floating aggregates in E5 medium (E8 medium without FGF2, TGF$\beta$-1, or insulin) +N2+B27+hFGF2 (4\nanogperml).
Four days later, we dissociated the aggregates with Accutase (Life Technologies Inc.) and plated them onto Matrigel coated plates in E5+N2+B27+hFGF2 (5ng/ml).
We cultured the cells for an additional 22 days and passaged when confluent.
After culture, we cryopreserved the cells at 12 million cells per vial.
We performed fluorescence-activated cell sorting (FACS) analysis and found that these NPCs were greater than 90\% SOX1$^+$ and Tub$\beta$III$^+$.

We thawed the NPCs and expanded them in E5+N2+B27+hFGF2 (5ng/ml) for five days before harvesting via Accutase treatment.
We then seeded roughly $10^5$ cells into one well of a 48 well Matrigel coated plate in E5+N2+B27 and started chemical treatment the same day (day 0).
We collected samples at indicated time points by lysing cells directly on plate with 150$\mu$l RLT lysis buffer (Qiagen).

\subsection*{RNA sequencing}
\label{sec:rnaseq}
After treatment with the compound, we purified total RNA from the NPCs using the RNeasy 96 Kit (Qiagen).
We prepared cDNA libraries, indexed with Illumina's TruSeq RNA Sequencing Kit, and sequenced on Illumina's HiSeq 2500, with 17-24 indexed samples per lane with 51 base-pair single reads.
We generated FASTQ files with CASAVA (v1.8.2) and mapped reads to the human transcriptome (RefGene v1.1.17) using Bowtie \cite{langmead2009ultrafast} (v0.12.8), allowing 2 mismatches and a maximum of 20 multiple hits.
To produce gene expression values in TPM, we used RSEM \cite{li2011rsem} (v1.2.1).

\subsection*{Classification experiments}
We used four common machine learning algorithms in our classification experiments: support vector machines (SVMs), logistic regression, random forests, and naive Bayes.
Traditional support vector machines (SVM) are binary classifiers that construct a maximum-margin decision boundary separating the two classes of training samples \cite{cortes1995support}.
Logistic regression is a binary classification algorithm that models the posterior probability of the response variable as a logistic function applied to a linear combination of the predictor variables and model coefficients \cite{logisticregression}.
Random forest classifiers produce class predictions by aggregating over an ensemble of decision trees \cite{breiman2001random}, each of which was fit using a bootstrap sample of the training dataset.
Naive Bayes is a probabilistic graphical model that makes the strong simplifying assumption that all predictor variables are conditionally independent of one another given the class label \cite{Mitchell}.

\begin{figure}[!ht]
  \centering
  \includegraphics[width=0.8\textwidth]{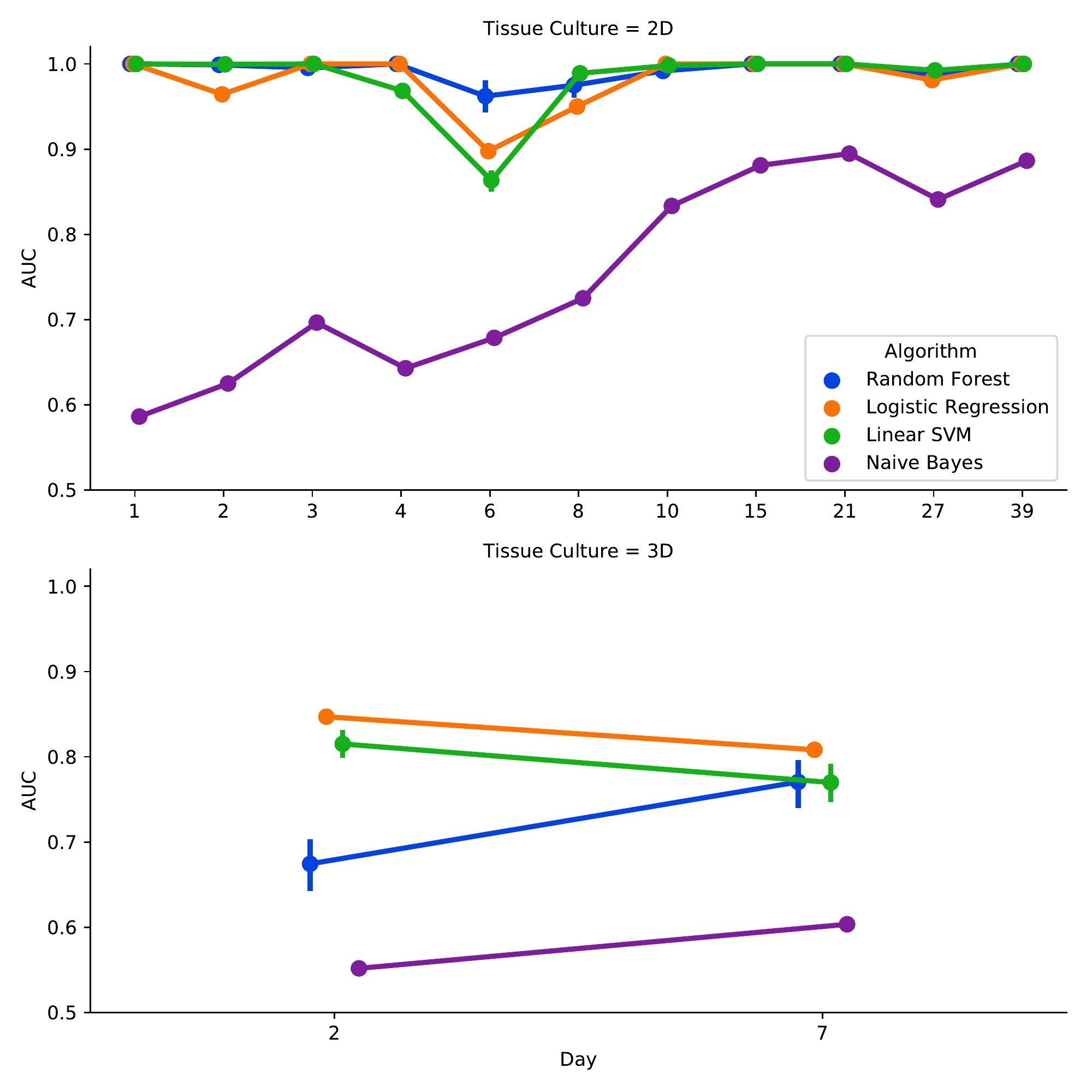}
  \caption{AUCs for single day train and test on the 2D (top) and 3D (bottom) tissue culture datasets. Error bars around points give a 95\% confidence interval from five replicate runs of each.}
  \label{fig:3d_and_2d_day_vs_day}
\end{figure}

We used the Scikit-learn \cite{scikit-learn} (v0.18.2) implementations of these algorithms, specifically the \texttt{SVC}, \texttt{LogisticRegression}, \texttt{RandomForestClassifier}, and \texttt{MultinomialNB} classes, respectively.
For SVMs, we used a linear kernel, probability estimates, and scaled gene expression values to [0.0, 1.0].
For logistic regression, we standardized the data, used L2 regularization, and used the dual formulation.
For both SVMs and logistic regression, we used internal cross-validation to select the C parameter from \{0.0001, 0.001, 0.01, 0.1, 1.0, 10.0, 100.0, 1000.0\}.
For random forests, we used the entropy splitting criterion and 100 trees.
We used the default settings for all other algorithm parameters.

To evaluate the predictive performance of these algorithms, we used the receiver operating characteristic (ROC) curve and the area under the curve (AUC).
We used this standard metric to get an overall sense of predictive performance without having to choose a single classification threshold for each model.
An AUC of 1.0 represents a perfect ordering of toxic and non-toxic compounds, whereas an AUC of 0.5 represents random guessing.

We used a standard leave-one-compound-out cross-validation for all of our experiments to avoid overly optimistic estimates of future predictive performance.
This means that we performed each experiment in 45 steps, corresponding with the 45 compounds in our dataset.
For each step, a single compound was held out of the training set, a model was trained on the remaining 44 compounds, and the model was used to make a prediction for the held out compound.
We then aggregated the predictions across all 45 compounds to make the final ROC curve evaluation.
Note that in experiments where we pooled multiple samples for each compound (e.g. samples from different time points), we held out all samples for the held out compound at once and averaged the predicted probabilities across samples to produce the final compound prediction.

\begin{figure}[!ht]
  \centering
  \includegraphics[width=0.8\textwidth]{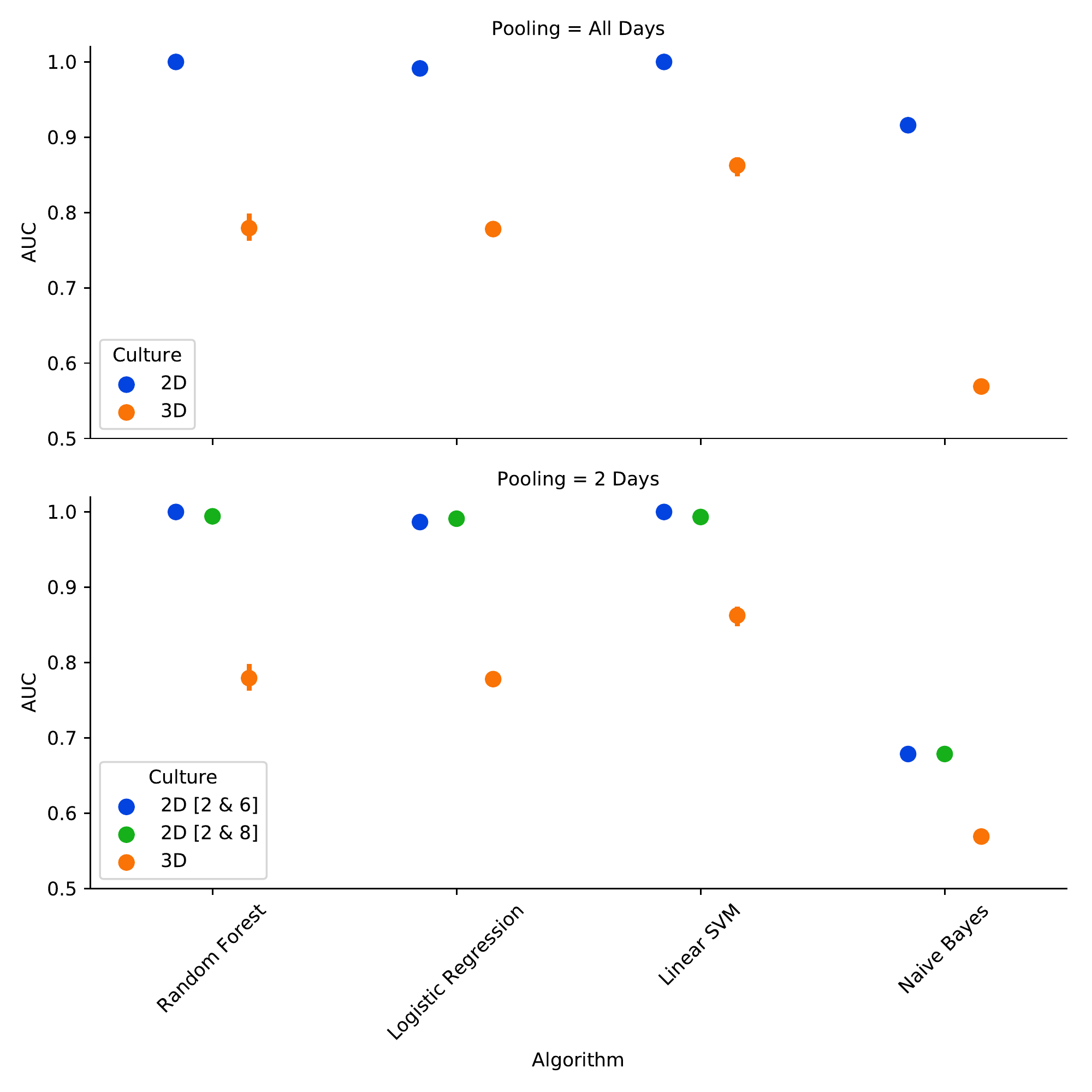}
  \caption{Comparison of AUCs between 2D and 3D tissue culture methods when all available exposure lengths are pooled for train and test (top), and when only two days of exposure are pooled for the 2D method (bottom). The 3D results are the same in both. Error bars around points give a 95\% confidence interval from five replicate runs of each experiment.}
  \label{fig:2d_vs_3d_boxplot_comparison}
\end{figure}

For our first classification experiment, we trained and tested our models using only one compound exposure length at a time.
The experiment shows there is variation in predictive performance over the extent of exposure lengths.
We ran five replicates of each to account for AUC variations between runs due to randomness in some of the algorithms.
Results for the 2D and 3D datasets can be found in Figure \ref{fig:3d_and_2d_day_vs_day}.

For our second classification experiment, we trained and tested our models by pooling samples from all exposure lengths together into one dataset.
Recall that we do not explicitly model the different exposure lengths, instead opting to simply include different exposure lengths as additional samples of the same compounds.
Again, we computed probability estimates for each compound by averaging the individual sample estimates.
As before, we ran five replicates of each.
Figure \ref{fig:2d_vs_3d_boxplot_comparison} (top) shows a comparison of AUCs for each of the algorithms between the 2D and 3D tissue culture methods. 

Because there are fewer exposure lengths in the 3D dataset, there is a chance that differences in accuracy between the 2D and 3D dataset are a result of having many more exposure lengths in the 2D dataset.
To account for this difference, we ran two smaller pooled experiments in the 2D dataset combining only two exposure lengths each.
The exposure lengths in the 3D dataset are two days and seven days, but we do not have a seven day exposure length in our 2D dataset.
Six and eight days of exposure are the closest analogous lengths in the 2D dataset.
Thus, we ran these two day experiments in two ways, once by pooling days two and six, and another by pooling days two and eight.
Again, we ran five replicates of each.
Figure \ref{fig:2d_vs_3d_boxplot_comparison} (bottom) shows the results.

\subsection*{Feature selection experiments}
We next trained the same models using all days pooled, but also applied feature selection.
We ran experiments on 19 different fixed sizes of selected gene sets: 1,000 genes down through 100 with a step size of 100, and through 10 genes with a step size of 10.
We chose these gene set sizes to demonstrate a wide range of model performance from large triple digit feature sets all the way down to small double digit sets.
We used the same feature selection algorithms for the 2D and 3D datasets, and all algorithms used the same set of selected features for their respective datasets.
We performed feature selection for each fold of cross-validation with the same data used for training the predictive models to avoid overly optimistic estimates of AUC.

We used three algorithms for our feature selection experiments: recursive feature elimination, filtering by mutual information, and sparse logistic regression.
Recursive feature elimination works by recursively training a linear model and eliminating the least important features in each step, until some stopping condition is met \cite{guyon2002rfe} (e.g. a desired feature set size).
Mutual information feature selection ranks all features by their mutual information with the class label and then filters to a specified feature set size \cite{yang1997mi}.
Sparse logistic regression works by applying L1, rather than L2, regularization to the model, driving many feature coefficients to 0 and thus leading to a smaller feature set \cite{friedman2001elements}.

Again, we used Scikit-learn \cite{scikit-learn} implementations for these algorithms, specifically the \texttt{RFE}, \texttt{SelectKBest}, and \texttt{LogisticRegression} classes, respectively.
For recursive feature elimination, we scaled expression values to [0.0, 1.0], used a linear SVM with C set to 1.0, and used a step size of 1\%.
For sparse logistic regression selection, we standardized the data, used L1 regularization, and set C to 1.0.
We used the default settings for all other parameters.

\begin{figure}[!h]
  \centering
  \includegraphics[width=0.8\textwidth]{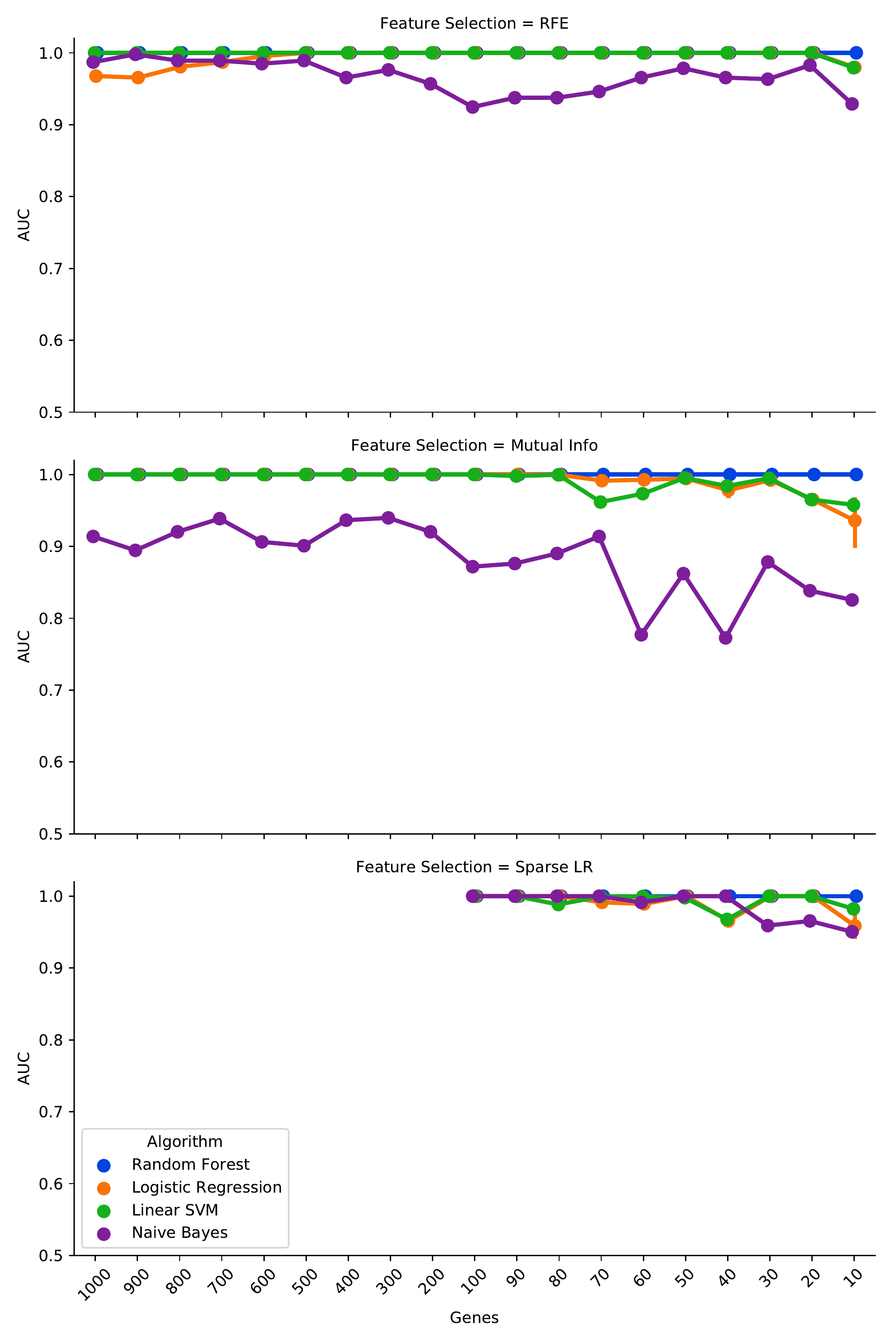}
  \caption{AUCs for models trained and tested with feature selection on all exposure lengths pooled for the 2D dataset. The horizontal axis is the selected gene set size. Error bars give a 95\% confidence interval from five replicate runs.}
  \label{fig:2d_hi_pooled_selk}
\end{figure}

\begin{figure}[!h]
  \centering
  \includegraphics[width=0.8\textwidth]{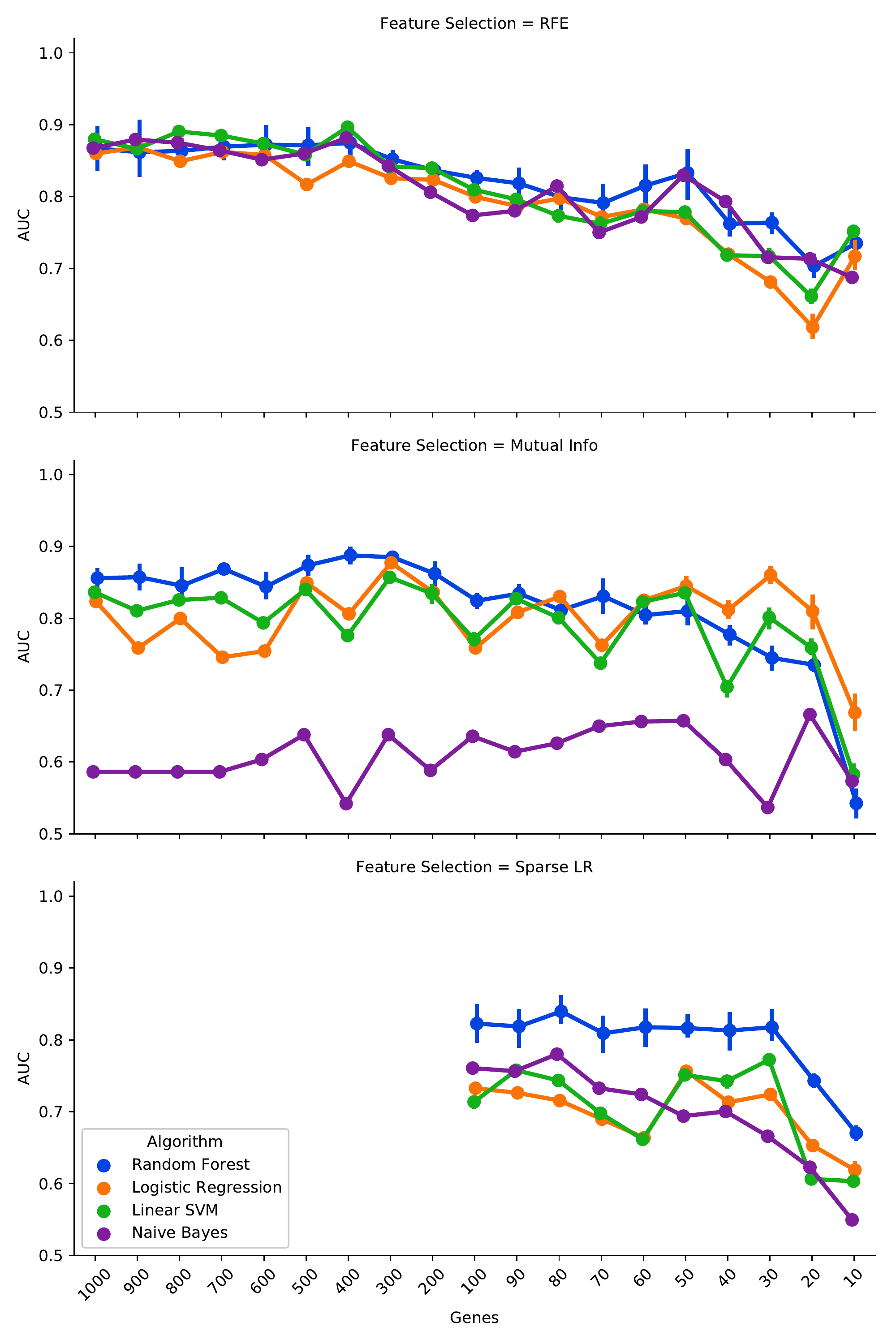}
  \caption{AUCs for models trained and tested with feature selection on all exposure lengths pooled for the 3D dataset. The horizontal axis is the selected gene set size. Error bars give a 95\% confidence interval from five replicate runs.}
  \label{fig:3d_pooled_selk}
\end{figure}

Note that the RFE and mutual information methods easily lend themselves to selection of exact numbers of genes, whereas spare logistic regression does not.
Thus, for sparse logistic regression, we first ran the sparse model and then selected the top K genes by the magnitude of their learned coefficients.
We skipped experiments with the sparse logistic regression method when K was larger than the number of non-zero learned coefficients.
As before, we ran five replicates of each experiment to account for AUC variations between runs.
Figures \ref{fig:2d_hi_pooled_selk} and \ref{fig:3d_pooled_selk} show the results from these experiments on the 2D and 3D datasets, respectively.

Because we performed feature selection for each fold of leave-one-compound-out cross-validation, each feature selection method was essentially performed 45 times for each gene set size.
To assess the variability of each feature selection method on these datasets, we report the number of unique genes selected across all folds combined for each method at a gene set size of 10 (see Table \ref{tab:selk_genes}).
Each feature selection method may thus pick a maximum of 450 genes, and a minimum of 10, across all folds.

\begin{table}[!ht]
\centering
\caption{Cross-validation unions of selected genes with K of 10.}
\label{tab:selk_genes}
\begin{tabular}{c c c c}
\toprule
Culture  & RFE & Mutual Info & Sparse LR \\
\midrule
2D & 99  & 25 & 37 \\
3D & 153 & 81 & 41 \\
\bottomrule
\end{tabular}
\end{table}

\section*{Results and Discussion}
Figure \ref{fig:3d_and_2d_day_vs_day} shows the single day-by-day prediction results for the 2D and 3D datasets.
Each point along the x-axis shows the predictive performance of a model that is trained and tested on samples at a single length of compound exposure in days.
Error bars around the points show a 95\% confidence interval from the five replicate runs of each algorithm and exposure length.
Recall that 1.0 is a perfect AUC, while 0.5 is equivalent to random guessing.
One commonality between the two culture methods is that naive Bayes does not appear to perform as well as the other three algorithms.
This is perhaps unsurprising due to the high dimensionality of the dataset, which skews naive Bayes' predictions more toward 0.0 and 1.0.
Predictive performance appears to improve overall at longer exposure lengths in the 2D dataset.
This same trend is not immediately apparent in the 3D dataset, but trends are difficult to support with only two exposure lengths.
The overall result, however, indicates that we can exceed the predictive performance of the 3D tissue model using the simpler 2D tissue model.

Figure \ref{fig:2d_vs_3d_boxplot_comparison} (top) shows a side-by-side comparison of the models on the 2D and 3D datasets when all available exposure lengths are pooled for training and testing.
Again, error bars around the points show a 95\% confidence interval from the five replicate runs of each.
All of the models have substantially better AUC on the 2D dataset than on the 3D dataset.
Naive Bayes again does not appear to perform as well as the other algorithms, but has substantially better accuracy on the 2D dataset once all of the samples are pooled.

Recall that the 2D dataset has several more compound exposure lengths than the 3D dataset.
Figure \ref{fig:2d_vs_3d_boxplot_comparison} (bottom) shows results on the 2D dataset when pooling only two days of compound exposure, which is analogous to the two available in the 3D dataset.
Again, naive Bayes does not perform as well as the other algorithms.
Still, in all cases, the 2D models perform better than the analogous experiments on the 3D dataset.
This suggests that the accuracy of the models on the 2D dataset is not simply a result of having a greater number of samples from a larger range of compound exposure lengths.

Finally, Figures \ref{fig:2d_hi_pooled_selk} and \ref{fig:3d_pooled_selk} show results from the same pooled models with the addition of three feature selection methods.
The horizontal axis gives the AUC at 19 decreasing gene set sizes, from 1,000 genes down to 10 genes.
The results show a common trend of accuracy degradation with smaller gene set sizes, but the trend is more pronounced in the 3D dataset.
For example, no method has an AUC above 0.8 in the 3D dataset at a gene set size of 10, whereas all but one method has an AUC above 0.9 in the 2D dataset.

In addition to better prediction, Table \ref{tab:selk_genes} shows that all of the feature selection methods generally selected more consistent sets of genes across folds on the 2D dataset, suggesting that the predictive signal may be less distributed in the 2D dataset than in the 3D dataset.
Note that there is no overlap between the 2D and 3D selected gene sets for any of the three methods, suggesting that the predictive signal is also distinct between 2D and 3D.


Overall, the 2D tissue model appears to produce more accurate and more consistent predictive models.
Two obvious potential explanations for this stand out.
First, the 2D tissue model, being composed of only one cell type, is less complex than the 3D model and thus produces a less variable signal.
Second, compound diffusion is likely more complete in the 2D tissue model than in 3D.
Both of these potential explanations could lead to stronger signals of gene expression perturbation for the machine learning algorithms to detect.

\section*{Conclusions}
Here, we present common machine learning models applied to the task of predicting developmental neurotoxicity of several compounds from gene expression data.
We compare the AUCs of these models between datasets collected from a 2D tissue culture approach, and a more complex 3D tissue culture approach.
We compare results from training models on single lengths of compound exposure, from multiple pooled lengths of exposure, and with the addition of feature selection.
Overall, our results show that the models trained on data collected from a simpler 2D tissue model are more accurate than those trained on data from a 3D model.
While a 3D tissue model is perhaps more likely to recapitulate relevant biology needed to fully understand toxicity mechanisms, a 2D tissue model is certainly a viable option and easier to produce efficiently \cite{chandrasekaran2017comparison}.
We would thus still recommend a 3D tissue model if the primary goal is to study biological mechanisms in depth, but our results here suggest that the 2D tissue model is an excellent choice for producing a broad toxicity assay.

Furthermore, our results show that models trained on the 2D data experience very little degradation in AUC under stringent feature selection, whereas the models trained on 3D data show extensive degradation.
Our results also demonstrate that the genes selected were more consistent across folds on the 2D data than on the 3D data; this is important because it suggests that we may be able to simplify the model even further by reducing the number of genes that need to be quantified to perhaps far fewer than 100 without loss of accuracy.
With a much smaller gene set, it may be possible to develop a similar assay using quantification methods that are still faster and cheaper than RNA-Seq.
We propose this direction of research for future work.

We further propose evaluating the use of 2D tissue models made from simpler cell types than NPCs to predict developmental neurotoxicity or toxicity in general.
NPCs require substantial experience to successfully differentiate and culture, whereas a tissue culture based on a cell type such as dermal fibroblasts may be more approachable.
While such a cell type may not be neural in nature, and the pattern of response would surely be different, the cell may still exhibit gene expression perturbations indicative of toxicity sufficient for the purposes of an assay.

Current models for toxicity screening are simply too slow and expensive to comprehensively test all new or less understood chemical exposures.
These models are certainly here to stay for the foreseeable future, but high-throughput screening methods, such as we have presented here, show a great deal of potential.
We hope and expect that both of these approaches may complement one another and accelerate findings by helping stakeholders choose which exposures to explore further and with what urgency.

\section*{Data Availability}
Our 2D tissue model is available through GEO Series accession number GSE126786, and the 3D tissue model data we use for comparison is available through GEO Series accession number GSE63935.
All code and processed expression data can be found on Github at \url{https://github.com/finnkuusisto/DevTox2D}.

\section*{Acknowledgements}
The authors acknowledge support from the National Institutes of Health (NIH) grant number UH3TR000506-05. The authors also thank Bao Kim Nguyen and Angela Elwell for technical assistance, John Steill for GEO submission assistance, and Marv and Mildred Conney for a grant to R. Stewart and J.A. Thomson.

\section*{Author Contributions}
F.K. and V.S.C. conceived computational experiments. F.K. conducted computational experiments, performed analysis and visualization, and wrote the original draft of the paper. Z.H. conducted cell culturing and differentiation. J.T., D.P., and R.S. conceived experiments, acquired funding, and administered the project. All authors reviewed the manuscript.

\section*{Competing Interests}
The authors declare no competing interests.

\bibliographystyle{unsrt}
\bibliography{devtox2d_paper_arxiv}

\begin{thebibliography}{10}

\bibitem{betts2010growing}
Kellyn~S Betts.
\newblock Growing knowledge: Using stem cells to study developmental
  neurotoxicity.
\newblock {\em Environmental Health Perspectives}, 118(10):A432, 2010.

\bibitem{rice2000critical}
Deborah Rice and Stan Barone~Jr.
\newblock Critical periods of vulnerability for the developing nervous system:
  Evidence from humans and animal models.
\newblock {\em Environmental Health Perspectives}, 108(Suppl 3):511, 2000.

\bibitem{grandjean2014neurobehavioural}
Philippe Grandjean and Philip~J Landrigan.
\newblock Neurobehavioural effects of developmental toxicity.
\newblock {\em The Lancet Neurology}, 13(3):330--338, 2014.

\bibitem{trasande2011reducing}
Leonardo Trasande and Yinghua Liu.
\newblock Reducing the staggering costs of environmental disease in children,
  estimated at \$76.6 billion in 2008.
\newblock {\em Health Affairs}, 30(5):863--870, 2011.

\bibitem{judson2014vitro}
Richard Judson, Keith Houck, Matt Martin, Thomas Knudsen, Russell~S Thomas,
  Nisha Sipes, Imran Shah, John Wambaugh, and Kevin Crofton.
\newblock In vitro and modelling approaches to risk assessment from the us
  environmental protection agency toxcast programme.
\newblock {\em Basic \& Clinical Pharmacology \& Toxicology}, 115(1):69--76,
  2014.

\bibitem{fabre2014organs}
Kristin~M Fabre, Christine Livingston, and Danilo~A Tagle.
\newblock Organs-on-chips (microphysiological systems): Tools to expedite
  efficacy and toxicity testing in human tissue.
\newblock {\em Experimental Biology and Medicine}, 239(9):1073--1077, 2014.

\bibitem{olson2000concordance}
Harry Olson, Graham Betton, Denise Robinson, Karluss Thomas, Alastair Monro,
  Gerald Kolaja, Patrick Lilly, James Sanders, Glenn Sipes, William Bracken,
  et~al.
\newblock Concordance of the toxicity of pharmaceuticals in humans and in
  animals.
\newblock {\em Regulatory Toxicology and Pharmacology}, 32(1):56--67, 2000.

\bibitem{hay2014clinical}
Michael Hay, David~W Thomas, John~L Craighead, Celia Economides, and Jesse
  Rosenthal.
\newblock Clinical development success rates for investigational drugs.
\newblock {\em Nature Biotechnology}, 32(1):40--51, 2014.

\bibitem{schwartz2015human}
Michael~P Schwartz, Zhonggang Hou, Nicholas~E Propson, Jue Zhang, Collin~J
  Engstrom, Vitor~Santos Costa, Peng Jiang, Bao~Kim Nguyen, Jennifer~M Bolin,
  William Daly, et~al.
\newblock Human pluripotent stem cell-derived neural constructs for predicting
  neural toxicity.
\newblock {\em Proceedings of the National Academy of Sciences},
  112(40):12516--12521, 2015.

\bibitem{chen2011chemically}
Guokai Chen, Daniel~R Gulbranson, Zhonggang Hou, Jennifer~M Bolin, Victor
  Ruotti, Mitchell~D Probasco, Kimberly Smuga-Otto, Sara~E Howden, Nicole~R
  Diol, Nicholas~E Propson, et~al.
\newblock Chemically defined conditions for human ipsc derivation and culture.
\newblock {\em Nature Methods}, 8(5):424--429, 2011.

\bibitem{beers2012passaging}
Jeanette Beers, Daniel~R Gulbranson, Nicole George, Lauren~I Siniscalchi,
  Jeffrey Jones, James~A Thomson, and Guokai Chen.
\newblock Passaging and colony expansion of human pluripotent stem cells by
  enzyme-free dissociation in chemically defined culture conditions.
\newblock {\em Nature Protocols}, 7(11):2029--2040, 2012.

\bibitem{langmead2009ultrafast}
Ben Langmead, Cole Trapnell, Mihai Pop, and Steven~L Salzberg.
\newblock Ultrafast and memory-efficient alignment of short dna sequences to
  the human genome.
\newblock {\em Genome Biology}, 10(3):R25, 2009.

\bibitem{li2011rsem}
Bo~Li and Colin~N Dewey.
\newblock {RSEM}: Accurate transcript quantification from rna-seq data with or
  without a reference genome.
\newblock {\em BMC Bioinformatics}, 12(1):323, 2011.

\bibitem{cortes1995support}
Corinna Cortes and Vladimir Vapnik.
\newblock Support-vector networks.
\newblock {\em Machine learning}, 20(3):273--297, 1995.

\bibitem{logisticregression}
D.~R. Cox.
\newblock The regression analysis of binary sequences.
\newblock {\em Journal of the Royal Statistical Society. Series B
  (Methodological)}, 20(2):215--242, 1958.

\bibitem{breiman2001random}
Leo Breiman.
\newblock Random forests.
\newblock {\em Machine learning}, 45(1):5--32, 2001.

\bibitem{Mitchell}
T~Mitchell.
\newblock {\em {Machine Learning}}.
\newblock McGraw-Hill, New York, 1997.

\bibitem{scikit-learn}
F.~Pedregosa, G.~Varoquaux, A.~Gramfort, V.~Michel, B.~Thirion, O.~Grisel,
  M.~Blondel, P.~Prettenhofer, R.~Weiss, V.~Dubourg, J.~Vanderplas, A.~Passos,
  D.~Cournapeau, M.~Brucher, M.~Perrot, and E.~Duchesnay.
\newblock Scikit-learn: Machine learning in python.
\newblock {\em Journal of Machine Learning Research}, 12:2825--2830, 2011.

\bibitem{guyon2002rfe}
Isabelle Guyon, Jason Weston, Stephen Barnhill, and Vladimir Vapnik.
\newblock Gene selection for cancer classification using support vector
  machines.
\newblock {\em Machine learning}, 46(1-3):389--422, 2002.

\bibitem{yang1997mi}
Yiming Yang and Jan~O Pedersen.
\newblock A comparative study on feature selection in text categorization.
\newblock In {\em ICML}, volume~97, pages 412--420, 1997.

\bibitem{friedman2001elements}
Jerome Friedman, Trevor Hastie, and Robert Tibshirani.
\newblock {\em The elements of statistical learning}.
\newblock Springer series in statistics, New York, 2001.

\bibitem{chandrasekaran2017comparison}
Abinaya Chandrasekaran, Hasan~X Avci, Anna Ochalek, Lone~N R{\"o}singh, Kinga
  Molnar, Lajos Laszlo, Tamas Bellak, Annamaria Teglasi, Krisztina Pesti, Arpad
  Mike, et~al.
\newblock Comparison of 2d and 3d neural induction methods for the generation
  of neural progenitor cells from human induced pluripotent stem cells.
\newblock {\em Stem Cell Research}, 25:139--151, 2017.

\end{thebibliography}

\end{document}